\documentclass[onecolumn,aps,superscriptaddress]{revtex4-2}
\usepackage[utf8]{inputenc}
\usepackage{epsf,graphicx}
\usepackage{multirow}
\usepackage{amsmath,gensymb,amssymb}
\usepackage{lipsum}
\usepackage{dcolumn}% Align table columns on decimal point
\usepackage{bm}% bold math
\usepackage{hyperref}
\usepackage{natbib}
\usepackage{color}

\begin{document}

\title{{\bf Acoustic turbulence: from the Zakharov-Sagdeev spectra to the Kadomtsev-Petviashvili spectrum}}

\author{E.A. Kochurin}
\email{kochurin@iep.uran.ru}
\affiliation{Institute of Electrophysics, Ural Branch of RAS, 620016, Yekaterinburg,  Russia}
\affiliation{Skolkovo Institute of Science and Technology, 121205, Moscow, Russia}
\author{E.A. Kuznetsov}
\affiliation{Skolkovo Institute of Science and Technology, 121205, Moscow, Russia}
\affiliation{Lebedev Physical Institute, RAS, 119991, Moscow, Russia}
\affiliation{Landau Institute for Theoretical Physics, RAS, Chernogolovka, 142432, Moscow
region, Russia}
\date{\today}

\begin{abstract}
This paper presents a brief review on theoretical and numerical works on three-dimensional acoustic turbulence both in a weakly nonlinear regime, when the amplitudes of sound waves are small, and in the case of strong nonlinearity. This review is based on the classical studies on weak acoustic turbulence by V.E. Zakharov (1965) and V.E. Zakharov and R.Z. Sagdeev (1970), on the one hand, and on the other hand, by B.B. Kadomtsev and V.I. Petviashvili (1972). Until recently, there were no convincing numerical experiments confirming one or the other point of view. In the works of the authors of this review in 2022 and 2024, strong arguments were found based on direct numerical modeling in favor of both theories. It is shown that the Zakharov-Sagdeev spectrum of weak  turbulence $\propto k^{-3/2}$ is realized not only for small positive dispersion of sound waves, but also in the case of complete absence of dispersion. The calculated turbulence spectra in the weakly nonlinear regime have anisotropic distribution: for small $k$, narrow cones (jets) are formed, broadening in the Fourier space. For weak dispersion, the jets are smoothed out, and the turbulence spectrum tends to be isotropic in the region of short wavelengths. In the absence of dispersion, the turbulence spectrum is a discrete set of jets subjected to diffraction divergence. For each individual jet, nonlinear effects are much weaker than diffraction ones, which prevents the formation of shock waves. Thus, the weakly turbulent Zakharov-Sagdeev spectrum is realized due to the smallness of nonlinear effects compared to dispersion or diffraction. As the pumping level increases in the non-dispersive regime, when nonlinear effects begin to dominate, shock waves are formed. As a result, acoustic turbulence passes into a strongly nonlinear state in the form of an ensemble of random shocks described by the Kadomtsev-Petviashvili spectrum $\propto k^{-2}$.

	\end{abstract}
	\maketitle	
	%\keywords{wave turbulence; capillary waves; Kolmogorov-Zakharov's spectra; direct numerical simulation; conformal mappings}

\section*{Introduction}
As is known, developed hydrodynamic turbulence at large Reynolds numbers, $Re\gg 1$,
in the inertial interval is an example of a system with strong nonlinearity, when its energy
coincides with the interaction Hamiltonian. Another classic
example is acoustic turbulence, which exhibits both strong and weak regimes depending on the ratio
of nonlinearity and linear wave characteristics. In this sense,
acoustic turbulence is much more diverse and richer than hydrodynamic
turbulence. When the nonlinear interaction of waves is small compared
to linear effects, a weak turbulence regime \cite{KZ-book} is realized, which can be studied on the basis of perturbation
theory using the random phase approximation. The theory of weak
turbulence \cite{KZ-book, naz-book} statistically describes ensembles of interacting
waves within the framework of the corresponding kinetic
equations \cite{KZ-book, naz-book}.
This theory assumes that each wave with its random phase
propagates almost freely for a long time and very rarely
undergoes small deformations due to nonlinear interaction with
other waves. To date, the theory of weak turbulence has found
many applications, ranging from ocean and plasma waves, waves in solid state
physics, in Bose-Einstein condensates and ending with
turbulence in both astrophysics and high-energy physics. Here we
will cite only some of the latest works on
this topic: \cite{Naz22, ShavitFalkovich,
FrahmShepelyansky, Semisalov, Galtier1, grav, Galtier, Shuryak}.
It should be noted that for waves on the surface of a liquid, both capillary
and gravitational (dispersive waves), the theory of weak turbulence
has been confirmed with very high accuracy \cite{FM}. However, for
acoustic turbulence, a sufficiently complete understanding has not been
yet. This is due to two different approaches to the study of acoustic
turbulence. The first works on this topic were carried out by V.E.~Zakharov in 1965 \cite{zakh65},
 and then in 1970 by V.E.~Zakharov and R.Z.~Sagdeev
\cite{zs-70} based on the theory of weak turbulence. As a
result, in the region of weak positive dispersion, a
scale-invariant turbulence spectrum $E(k)\propto k^{-3/2}$ was found,
called the Zakharov-Sagdeev spectrum. This spectrum
was an exact solution in the long-wave limit
of the stationary kinetic equation for waves, proportional
to the square root of the energy flux $\epsilon$ into the region of short
wavelengths.
In 1972, B.B. Kadomtsev and V.I. Petviashvili \cite{KP}
proposed their spectrum, which is now called the Kadomtsev-Petviashvili spectrum, which has a different dependence $E(k)\propto
k^{-2}$. According to \cite{KP}, acoustic turbulence was considered
as an ensemble of random shock waves. Shock waves arise as a result of the breaking of sound waves of finite amplitude.
The $k^{-2}$ dependence is due to density jumps. Until recently, there was not a single direct numerical
experiment in which one or the other spectrum was observed. In this review, based on numerical experiments
\cite{KochurinKuznetsov1}, \cite{KochurinKuznetsov2}, it will be shown what are the criteria for the emergence of the Zakharov-Sagdeev and Kadomtsev-Petviashvili spectra. In both cases, for small but finite amplitudes, the main nonlinear process is the three-wave interaction, which is resonant when the synchronism conditions are met
\begin{equation}  \label{3wave}
\omega(\mathbf{k})=\omega(\mathbf{k_1})+ \omega(\mathbf{k_2}), \quad
\mathbf{k }=\mathbf{k_1}+\mathbf{k_2}
\end{equation}
where $\omega=\omega(\mathbf{k})$ is the dispersion law of linear
waves. If $\omega(\mathbf{k})=kc_s$, where $c_s$ is the speed of sound, then
these relations are obviously satisfied only for parallel
wave vectors $\mathbf{k_i}$. Thus, a triplet
of interacting waves forms a ray in $k$-space. Obviously,
the transition to a coordinate system moving with the velocity $c_s$
along a separate ray makes this wave system strongly nonlinear. In
one-dimensional gas dynamics, this nonlinearity leads to the breaking
of acoustic waves in accordance with the famous Riemann solution. In
the multidimensional case, the transition to a moving coordinate system with the
velocity $c_s$ is possible only for one given ray; for all
other rays propagating at certain angles, such a transition is
not correct. A set of continuously distributed beams with close angles of
wave propagation forms an acoustic beam, which, as is known, is subject to diffraction
in the transverse direction. For any small but finite transverse beam width, the phenomenon
of diffraction divergence will inevitably arise.
This effect was first discussed in the original paper by Zakharov and Sagdeev \cite{zs-70};
at almost the same time, similar ideas were developed in the work of \cite{Newell} (see also \cite{AT}).
From these considerations it follows that the regime of weak turbulence for acoustic waves can
be realized not only in the case of weak dispersion, but also in a dispersion-free situation due
to the diffraction divergence of the wave beam.

It should be noted that the resonance conditions (\ref{3wave}) are also satisfied for acoustic waves with weak positive dispersion, when
\begin{equation}
\omega =kc_{s}(1+a^{2}k^{2}),\quad(a^{2}k^{2}\ll 1),  \label{disp1}
\end{equation}
where $a$ is the dispersion parameter, which has the dimension of length. In
the weakly dispersive case, the interacting waves form cones with small angles
 $\sim (ak)^{2}$ instead of rays. In the case of small nonlinear
effects compared to the dispersion of the waves, a weakly
turbulent regime of the evolution of sound waves can be realized. For this case, the exact
isotropic solution of the kinetic equation for three-dimensional weak
acoustic turbulence, obtained in \cite{zakh65, zs-70}, has the form
\begin{equation}
E(k)=C_{KZ}\left( \rho c_{s}\epsilon \right) ^{1/2}k^{-3/2},  \label{ZS}
\end{equation}
where $C_{KZ}$ is the dimensionless Kolmogorov-Zakharov constant, $\rho $ is the equilibrium gas density.
The power-law dependence on $\epsilon $ in \eqref{ZS} with the exponent $1/2$ corresponds to
the resonant three-wave interactions \eqref{3wave}. In the long-wave limit, such a spectrum
is local and does not depend on the dispersion parameter
of waves $a$ \cite{zakh65, zs-70}. However, in two-dimensional geometry, the spectrum
of weak acoustic turbulence explicitly contains a dependence
on $a$ \cite{Naz22}: $E(k)= C (\epsilon a c_s)^{1/2}/k$, where $C$ is a dimensionless constant.
 It should be noted that this dependence of the spectrum on $k$ was given in \cite{DNPZ} without derivation.
As for the one-dimensional case, the theory of weak sound turbulence turns out to be inapplicable,
as was shown by Benney and Saffman \cite{BenneySaffman}.

The existence of the Zakharov-Sagdeev spectrum (\ref{ZS}) in the inertial interval, as a Kolmogorov-type spectrum,
was confirmed in a number of works \cite{kin1, kin2, kin3} in numerical solution of the kinetic equation
 for waves in the presence of long-wave pumping and high-frequency attenuation. It is important to note that
numerical study of weak sound turbulence within the framework of the kinetic
equation for waves in the three-dimensional case was carried out only for
isotropic distributions.

With an increase in energy pumping, when the effects of nonlinear interaction of waves become
comparable with linear dispersion, various nonlinear coherent structures arise:
solitons, collapses, shock waves, etc. (see,
for example, \cite{ZakharovKuznetsov2012} and \cite{Kuznetsov2022}). In
this case, the behavior of the system is determined by the interaction
of coherent structures with an ensemble of incoherent chaotic waves.

In the non-dispersive case, when nonlinear effects prevail over diffraction,
the breaking of acoustic waves leads to the formation of propagating density discontinuities, which are the shock waves.
In this regime, according to Kadomtsev and Petviashvili \cite{KP}, acoustic turbulence
 is an ensemble of randomly distributed shocks. In this case, the spectrum of acoustic turbulence has the form:
\begin{equation}
E(k)= \frac{2S_d c_s^2 \langle \delta \rho^2\rangle}{\pi \rho}k^{-2},
\label{KP}
\end{equation}
where $S_d$ is the density of shock waves per unit length, $\langle \delta
\rho^2\rangle$ is the mean square of density jumps at discontinuities
(see, for example, \cite{Kuz04}). Note that the exponent
of the power spectrum \eqref{KP} does not depend on the dimension of space.
In the one-dimensional case, the expression (\ref{KP}) is nothing more than the spectrum
of turbulence, first found by Burgers \cite{Burgers}.
Thus, at the present time, there are two different
approaches to the description of acoustic non-dispersive turbulence,
predicting different spectra for the same phenomenon.

The main objective of this review is to answer the fundamental question about the
reason for the different behavior of the acoustic
turbulence spectra (\ref{ZS}) and (\ref{KP}). The paper presents the results of
direct numerical simulation of sound wave turbulence in
three-dimensional geometry \cite{KochurinKuznetsov1,KochurinKuznetsov2},
which indicate the possibility of realizing both the weak turbulence Zakharov-Sagdeev  spectrum
 \eqref{ZS} and the spectrum of Kadomtsev-Petviashvili (\ref{KP}) for three-dimensional acoustic waves in
the regime of strong sound turbulence. The transition between these regimes
is determined by the level of nonlinearity of the system. In the works of
authors \cite{KochurinKuznetsov1,KochurinKuznetsov2} in 2022 and 2024, compelling criteria were found in favor of both
theories. It is shown that the Zakharov-Sagdeev weak turbulence spectrum 
 $\propto k^{-3/2}$ is realized not only at
small positive dispersion of sound waves, but also in the case of complete
absence of dispersion. The calculated turbulence spectra in the weakly
nonlinear regime have an anisotropic distribution: in the region
of small $k$, narrow cones (jets) are formed, broadening with increasing $k$ in
Fourier space. In the case of weak dispersion, the jets are smoothed out, and the spectrum
of turbulence tends to be isotropic in the region of short wavelengths. In the absence of dispersion,
the turbulence spectrum is a discrete set of jets subjected to diffraction divergence. It was found that for each individual jet, nonlinear
effects are much weaker than diffraction ones, which prevents the formation of shock waves. Thus, the weak turbulence Zakharov-Sagdeev
spectra are realized due to the smallness of nonlinear effects in
comparison with dispersion or diffraction. With an increase in the pumping level in the
dispersionless regime, when nonlinear effects begin to prevail,
shock waves (density discontinuities propagating in space) are formed. In a result,
acoustic turbulence passes into a strongly nonlinear
state in the form of an ensemble of random shocks described by the Kadomtsev-Petviashvili
spectrum $\propto k^{-2}$.

The plan of the article is as follows. In the second section, we will consider in detail weak
turbulence of sound waves using the kinetic equation and indicate the criteria for its applicability.
In the third section, following the works of Zakharov and Sagdeev \cite{zakh65, zs-70}, we will show
how the Zakharov-Sagdeev spectrum is analytically found based on the solution of the kinetic equation for waves.
The next section presents the numerical modeling scheme and its parameters.
The fourth section presents the results of direct numerical modeling in both weak and strong turbulence modes.
In addition to the turbulence spectra and their distributions, results on the statistics of turbulent states are presented. For
weak turbulence, the amplitude distribution is close to Gaussian, and in
the case of strong turbulence, the probability density distribution
has power tails, indicating intermittency caused by shock waves.

This work is based on the lecture of E.A.
Kuznetsov \cite{Kuznetsov-2024} read at the school "Nonlinear Waves - 2024".

\section{Kinetic equation}

In this review, following the works of Zakharov and Sagdeev \cite{zakh65, zs-70}, we will describe three-dimensional acoustic
turbulence within the framework of the nonlinear string equation,
often called the Boussinesq equation. This equation is written for a scalar quantity $u(\mathbf{r},t)$, depending on three spatial
coordinates $\mathbf{r}=\{x,y,z\}$ and time $t$:
\begin{equation}
u_{tt}=\Delta u-2a^{2}\Delta ^{2}u+\Delta (u^{2}),  \label{eq0}
\end{equation}
where $\Delta$ is the Laplace operator. The dimensionless
quantity $u(\mathbf{r},t)$ in this equation has the meaning of a fluctuation
of density, which can be seen if we rewrite this equation as a
system of two equations \begin{eqnarray}
u_{t} + \Delta \phi &=& 0,  \label{density} \\
\phi _{t}+ u + u^{2} &=& 2 a^{2}\Delta u,  \label{phase}
\end{eqnarray}
where $\phi$ is the hydrodynamic
velocity potential $\mathbf{v}=\nabla \phi$. In these equations,
the speed of sound $c_s$ and the average density $\rho$ are set
to $1$. The parameter $a$ is the dispersion length.
The system of equations (\ref{density}) and (\ref{phase}) differs from the traditional system of gas dynamics, for
which the left-hand side of (\ref{density}) additionally
contains the term $\mbox{div} u \nabla \phi$, and in the equation (\ref{phase}) there is
the term $(\nabla \phi)^2/2$. As will be shown below, in the
long-wave limit $ka\ll 1$ when describing a three-wave
nonlinear interaction at moderate amplitudes,
the wave vectors $\mathbf{k_i}$ of the interacting waves are practically
parallel to each other. In this case, the three-wave matrix element
has the same dependence
on the vectors $\mathbf{k_i}$ and differs from the correct
``gas-dynamic'' system only by a factor (see review
\cite{ZakharovKuznetsov1997}).

Note that equation (\ref{eq0}) in the one-dimensional case refers to
equations integrable by the inverse scattering method
\cite{Zakharov1973}. In three-dimensional geometry, this model was first used to study weak
acoustic turbulence by
Zakharov in his work \cite{zakh65}. In the linear approximation,
equation (\ref{eq0}) has the dispersion law
\begin{equation}
\omega ^{2}=k^{2}+2a^{2}k^{4},\qquad k=|\mathbf{k}|,  \label{disp}
\end{equation}
coinciding
with the Bogolyubov spectrum for oscillations of the condensate of a weakly nonideal
Bose gas. In the case of weak dispersion $ka\ll 1$, this dispersion law
transforms into (\ref{disp1}), which at $ka\ll 1$
transforms into \eqref{disp1}.

The equations \eqref{density}, \eqref{phase}
relate to Hamiltonian systems and can be represented as
\begin{equation}
u_{t}=\frac{\delta H}{\delta \phi },\qquad \phi _{t}=-\frac{\delta H}{\delta
u},  \label{ham}
\end{equation}
where the Hamiltonian $H$ is written as
\begin{equation*}
H=\frac{1}{2}\int \left[ \left( \nabla \phi \right) ^{2}+u^{2}\right] d
\mathbf{r}+\int a^{2}(\nabla u)^{2}d\mathbf{r}+\frac{1}{3}\int
u^{3}d\mathbf{\ r}\equiv
\end{equation*}
\begin{equation}
\equiv H_{1}+H_{2}+H_{3}.  \label{ham1}
\end{equation}
In the Hamiltonian (\ref{ham1}) three terms are distinguished.
The first $H_{1}$ is the sum of the kinetic and potential
energy of linear non-dispersive
waves. The second term $H_{2}$ is responsible for the dispersion part
of the energy, and $H_{3}$ is responsible for the nonlinear interaction of the waves.

Carrying out further Fourier transform with respect to spatial variables and
introducing normal variables $a_{k}$ and $a_{k}^{\ast }$, \begin{equation*}
u_{k}=\left( \frac{k^{2}}{2\omega _{k}}\right) ^{1/2}(a_{k}+a_{-k}^{\ast }),
\end{equation*}\begin{equation*}
\phi _{k}=-i\left( \frac{\omega _{k}}{2k^{2}}\right)
^{1/2}(a_{k}-a_{-k}^{\ast }),
\end{equation*}the equations \eqref{ham} are reduced to the standard
form \cite{KZ-book, naz-book,ZakharovKuznetsov1997}: \begin{equation}
\frac{\partial a_{k}}{\partial t}=-i\frac{\delta H}{\delta a_{k}^{\ast }},
\label{normal}
\end{equation}
Where
\begin{equation*}
H=\int \omega _{k}|a_{k}|^{2}d\mathbf{k}+\int V_{k_{1}k_{2}k_{3}}\left(
a_{k_{1}}^{\ast }a_{k_{2}}a_{k_{3}}+a_{k_{1}}a_{k_{2}}^{\ast
}a_{k_{3}}^{\ast }\right) \delta \left(
\mathbf{k}_{1}-\mathbf{k}_{2}-\mathbf{k}_{3}\right) d\mathbf{k}_{1}d\mathbf{k}_{2}d\mathbf{k}_{3}.
\end{equation*}
In the nonlinear Hamiltonian $H_{3}$ we left
only one resonance term, corresponding to the decay processes
of the waves \eqref{3wave}. In the long-wave
approximation
$ka\ll 1$ we will take into account the dispersion only in the quadratic part of $H$ and
neglect
it in the matrix element
\begin{equation}
V_{k_{1}k_{2}k_{3}}=\frac{1}{8\pi ^{3/2}}\left( k_{1}k_{2}k_{3}\right)
^{1/2}. \label{matrix}
\end{equation}
This expression coincides with the
matrix element for acoustic waves up to a constant
\cite{KZ-book, zakh65, ZakharovKuznetsov1997}. As a result, the kinetic
equation for the pair correlator $n_{k}$,
defined as $\langle a_{k}^{\ast }a_{k_{1}}\rangle
=n_{k}\delta (\mathbf{k}-\mathbf{k}_{1})$, in the weak
turbulence approximation is written as follows:
\begin{equation}
\frac{\partial n_{k}}{\partial t}=2\pi \int
d\mathbf{k}_{1}d\mathbf{k}_{2}\left( T_{kk_{1}k_{2}}-T_{k_{1}kk_{2}}-T_{k_{2}kk_{1}}\right) ,
\label{kin}
\end{equation}
where
\begin{equation*}
T_{kk_{1}k_{2}}=|V_{kk_{1}k_{2}}|^{2}(n_{k_{1}}n_{k_{2}}-n_{k}n_{k_{2}}-n_{k}n_{k_{1}})\delta \left(
\mathbf{k}-\mathbf{k}_{1}-\mathbf{k}_{2}\right) \delta \left( \omega
_{k}-\omega _{k_{1}}-\omega _{k_{2}}\right) .
\end{equation*}
In this equation, $n_{k}$, the number of waves, is the
classical analogue of occupation numbers. This kinetic
equation can be obtained from the quantum kinetic equation in the
limit of large occupation numbers. In the quantum case, the kinetic
equation was first derived by Peierls \cite{Peierls}, and then
was used in the work of Landau and Rumer \cite{LandauRumer} to
find the attenuation of acoustic waves in a solid. Note
that the kinetic equation \eqref{kin}, in accordance with
Fermi's rule, contains in the integral the factor $2\pi |V_{kk_{1}k_{2}}|^{2}$,
multiplied by two delta functions expressing the laws of conservation of momentum and
energy \eqref{3wave}. As for the derivation of this equation for
classical wave fields, it can be obtained using the random phase approximation. To do this,
we first obtain an equation for the pair correlator $n_{k}$, which
contains a triple correlator $J_{kk_{1}k_{2}}$, characterizing
the correlation of three interacting waves. In turn, the equation for
the triple correlator contains a fourth-order correlator. In
the equation for $J_{kk_{1}k_{2}}$, we should neglect the derivative of $J_{kk_{1}k_{2}}$ with respect
to time, and the quadruple correlator should be represented as a sum
of products of pair correlators, neglecting the quadruple
cumulant. As a result, for the pair correlator $n_{k}$, we obtain a
closed kinetic equation \eqref{kin}. Such a description of wave
turbulence
is valid if the dispersion effects are large compared to the nonlinear
interaction (for details, see, for example, \cite{KZ-book, naz-book}).
In the case of acoustic waves described by the equations \eqref{density} and \eqref{phase}, this means that the nonlinear
effects for which the Hamiltonian $H_{3}$ is responsible are small compared to $H_{2}$.

In order to calculate the turbulence energy spectrum $E(k)$, it is necessary to first find the distribution
of $n_{k}$ as a solution to the equation \eqref{kin}, and then average the expression over the angles (see \cite{landau}):
\begin{equation*}
E(k)=\frac{1}{(2\pi )^{3}}k^{2}\omega _{k}\int n(\mathbf{k})d\Omega ,
\end{equation*}
where $d\Omega $ is the element of the solid angle.

\section{Zakharov-Sagdeev's spectrum}
At the beginning of this section we will show how the Zakharov-Sagdeev  spectrum \eqref{ZS}
 is obtained from the kinetic equation and then we will find
the Kolmogorov-Zakharov constant included in the expression for the spectrum of \eqref{ZS}.

We will consider stationary spherically symmetric solutions in the
long-wave limit. It is in this case that we can obtain
a scale-invariant distribution $n_{k}=Ak^{x}$, where $A$ is a
constant and ${x}$ is the power to be determined. As was first
noted by Zakharov \cite{zakh65}, in the kinetic equation
\eqref{kin} in the isotropic case in $\omega
_{k}$ at $k\rightarrow 0$ the dispersion can be neglected, despite
the singularity caused by the product of two $\delta
$-functions. It turns out that after averaging over the angles $\delta $ -
functions of the wave vectors $\mathbf{k_{i}}$,
\begin{equation} \label{mean}
\left\langle \delta \left( \mathbf{k}-\mathbf{k}_{1}-\mathbf{k}_{2}\right)
\right\rangle =2\pi /(kk_{1}k_{2}),
\end{equation}
the singularity in equation (\ref{kin}) becomes
integrable. In this case, the equation (\ref{kin}) in the stationary
spherically isotropic case can be represented as
\begin{equation} \label{stateq}
\int dk_{1}dk_{2}\left(
W_{kk_{1}k_{2}}-W_{k_{1}kk_{2}}-W_{k_{2}kk_{1}}\right) =0,
\end{equation}
where
\begin{equation*}
W_{kk_{1}k_{2}}=k^{2}k_{1}^{2}k_{2}^{2}(n_{k_{1}}n_{k_{2}}-n_{k}n_{k_{2}}-n_{k}n_{k_{1}})\delta \left( k-k_{1}-k_{2}\right) .
\end{equation*}
With the power dependence $n_{k}=Ak^{x}$
the function $W_{kk_{1}k_{2}}$ turns out to be homogeneous with respect to
its arguments of degree $z=2x+5$. This allows in (\ref{stateq})
to perform the Zakharov transformations in the second and third integrals
\begin{eqnarray*}
k_{1} &=&k\frac{k}{k^{\prime }},k=k^{\prime }\frac{k}{k^{\prime
}},k_{2}=k^{\prime \prime }\frac{k}{k^{\prime }}; \\
k_{2} &=&k\frac{k}{k^{\prime \prime }},k=k^{\prime \prime
}\frac{k}{k^{\prime \prime }},k_{1}=k^{\prime }\frac{k}{k^{\prime \prime }}.
\end{eqnarray*}
As a result, the integration domains in the second
and third terms of the equation (\ref{stateq}) coincide with the integration domain
of the first integral. Secondly, due to the homogeneity of $W_{kk_{1}k_{2}}$, the stationary equation (\ref{stateq}) is transformed to the form:
\begin{equation} \label{result}
\int dk_{1}dk_{2}
W_{kk_{1}k_{2}}\left[ 1-\left(\frac{k}{k_1}\right)^y-\left(\frac{k}{k_2}\right)^y\right] =0,
\end{equation}
where $y=2x+8$. Due to the presence of $W_{kk_{1}k_{2}}$ $\delta (k-k_{1}-k_{2})$, the expression in square brackets (\ref{result}) vanishes at $y=-1$, which
corresponds to the Zakharov-Sagdeev spectrum. In this case, $W_{kk_{1}k_{2}}$ identically vanishes at $n_k\propto k^{-1}$, which corresponds to the thermodynamically equilibrium Rayleigh-Jeans spectrum.
	As was first shown in \cite{zakh65, zs-70}, the Zakharov-Sagdeev spectrum
is a Kolmogorov-type spectrum corresponding to a constant energy flow
$\epsilon$ raised to the power of $1/2$. To establish this fact, one must use the non-stationary equation (\ref{kin}).
Let's multiply \eqref{kin} by $4\pi k^{3}\frac{1}{(2\pi )^{3}}$, hence, taking into account (\ref{mean}) we have
\begin{equation}
\frac{\partial E(k)}{\partial t}=\frac{1}{32\pi ^{3}}k\int
dk_{1}dk_{2}\left( W_{kk_{1}k_{2}}-W_{k_{1}kk_{2}}-W_{k_{2}kk_{1}}\right)
\label{E(k)}.
\end{equation}
Since $\int
E(k)dk=const$, equation (\ref{E(k)}) can be rewritten as a conservation
law:
\begin{equation*}
\frac{\partial E(k)}{\partial t}+\frac{\partial \epsilon}{\partial k}=0
\end{equation*}
where $\epsilon$ is the energy flux depending on $k$.
This value is found after
integrating
the left-hand side (\ref{E(k)}) over $k$
\begin{equation}
\frac{\partial \epsilon}{\partial k}=-\frac{k}{32\pi ^{3}}\int dk_{1}dk_{2}\left(
W_{kk_{1}k_{2}}-W_{k_{1}kk_{2}}-W_{k_{2}kk_{1}}\right) . \label{derivative}
\end{equation}
Substituting $n_{k}=Ak^{x}$ into this relation allows
to represent
$\frac{\partial \epsilon}{\partial k}$ as a power function
\begin{equation}
\frac{\partial \epsilon}{\partial k}=k^{y}I;\qquad y=2x+8, \label{flux-y}
\end{equation}
where $I$ is a constant depending on $y$.
As a result,
we obtain the relation
\begin{equation}
\epsilon=\frac{k^{y+1}}{y+1}I. \label{flux}
\end{equation}
Note that $\epsilon$ does not depend on $k$ for $y=-1$, where $I=0$ by (\ref{stateq}),
which corresponds to the exact stationary solution
of the equation (\ref{kin}) in the form of the Zakharov-Sagdeev spectrum:
\begin{equation*}
E(k)\propto k^{-3/2}.
\end{equation*}
The dependence
of $I$ on $x$ is found using the Zakharov transformations of the second and third
terms in (\ref{derivative}).
Note that the stationary spectrum of weak turbulence
corresponds to the parameters:
\begin{equation*}
y=-1,\,\,\mbox{or}\,\,x=-9/2.
\end{equation*}
After integrating the equation (\ref{flux}) with respect to
$k_{2}$, we obtain the equation
\begin{equation*}
\frac{\partial \epsilon}{\partial k}=-\frac{A^{2}}{32\pi
^{3}}k\int_{0}^{k}k^{2}k_{1}^{2}(k-k_{1})^{2}\left[ 1-\left(
\frac{k}{k_{1}}\right) ^{y}-\left( \frac{k}{k-k_{1}}\right) ^{y}\right] \left[
kk_{1}(k-k_{1})\right] ^{x}\cdot
\end{equation*}
\begin{equation*}
\cdot \left[ k^{-x}-k_{1}^{-x}-(k-k_{1})^{-x}\right] dk_{1}.
\end{equation*}
It is convenient to introduce dimensionless
parameter $\xi =k_{1}/k$ ($0\leq \xi \leq 1$):
\begin{equation*}
\frac{\partial \epsilon}{\partial k}=-\frac{A^{2}}{32\pi ^{3}}k^{y}\int_{0}^{1}d\xi
\left[ 1-\left( \frac{1}{\xi }\right) ^{y}-\left( \frac{1}{1-\xi }\right)
^{y}\right] \left[ \xi (1-\xi )\right] ^{x+2}\left[ 1-\xi ^{-x}-(1-\xi
)^{-x}\right]
\end{equation*}
Where
\begin{equation*}
I(y)=-\frac{A^{2}}{32\pi ^{3}}\int_{0}^{1}d\xi \left[ 1-\left( \frac{1}{\xi
}\right) ^{y}-\left( \frac{1}{1-\xi }\right) ^{y}\right] \left[ \xi (1-\xi
)\right] ^{x+2}\left[ 1-\xi ^{-x}-(1-\xi )^{-x}\right] .
\end{equation*}
Substituting further into (\ref{flux}) $y=-1+\alpha $ and finding
the limit $\alpha \rightarrow 0$, we obtain the following expression for
the energy flux $\epsilon$:
\begin{eqnarray}
\epsilon &=&\frac{A^{2}}{32\pi ^{3}}\Lambda , \label{square} \\
\Lambda &=&\int_{0}^{1}d\xi \left[ \xi \ln \left( \frac{1}{\xi }\right)
+(1-\xi )\ln \left( \frac{1}{1-\xi }\right) \right] \left[ \xi (1-\xi
)\right] ^{-5/2}\left[ 1-\xi ^{9/2}-(1-\xi )^{9/2}\right] . \notag
\end{eqnarray}
Here $\Lambda $ is a convergent positive-sign
integral $\Lambda >0$, providing an energy flux
directed to the short-wave region. Note also that
the convergence of this integral means the locality of the spectrum.
Thus, the constant $A$ in the equation (\ref{flux}) takes the form
\begin{equation*}
A=4\pi \left( \frac{2\pi \epsilon }{\Lambda }\right) ^{1/2}.
\end{equation*}
Therefore, the Zakharov-Sagdeev spectrum can be written as
\begin{equation*}
E(k)=\left( \frac{8\epsilon }{\pi \Lambda }\right) ^{1/2}k^{-3/2}?
\end{equation*}
where $C_{KZ}$ is the Kolmogorov-Zakharov constant
is expressed as follows
\begin{equation*}
C_{KZ}=\left( \frac{8}{\pi \Lambda }\right) ^{1/2},
\end{equation*}
while the integral $\Lambda $ is calculated explicitly:
$\Lambda =\frac{32}{3}(\pi -1+\ln 16)$.
Thus, in the isotropic
case, for the Kolmogorov-Zakharov constant we have
\begin{equation}
C_{KZ}=\left[ \frac{3}{4\pi (\pi -1+\ln 16)}\right] ^{1/2}\approx 0.22.
\label{KZaa}
\end{equation}
Note that a very close value of the Kolmogorov-Zakharov
constant was recently obtained in \cite{Galtier1} for fast
magnetoacoustic waves at small beta values, which are also waves
with weak dispersion.

\section{Numerical model and parameters}
To model stationary turbulence, we introduce into the \eqref{ham} equations the terms responsible for pumping and dissipating
energy:
\begin{equation}
u_{t}=-\Delta \phi +\mathcal{F}(\mathbf{k},t)-\gamma _{k}u, \label{eq1}
\end{equation}
\begin{equation}
\phi _{t}=-u+2a^{2} \Delta u-u^{2}, \label{eq2}
\end{equation}
where the viscosity operator $\gamma _{k}
$ takes nonzero values at $k\geq k_d$, and the random external
force $\mathcal{F}(\mathbf{k},t)$ is localized in the region of large
scales $k\leq k_2$ with a maximum $F_0$
achieved at $k_1$ ($k_d\gg k_2$). The dissipation energy flux
$\epsilon_N(t)$ is calculated as
\begin{equation} \label{Ds}
\epsilon_N(t)=L^{-3}\int\limits_{L^3} u \hat {f}^{-1}(\gamma_k u_k)
d\mathbf{r},
\end{equation}
where $\hat{f}^{-1}$ is the inverse Fourier transform and $L$ is the size
of the system. In the steady state, the energy flux is defined as
$\epsilon=\langle\epsilon_N\rangle=T^{-1}\int_0^T \epsilon_N(t) dt$.

Numerical integration of the system of equations (\ref{eq1}) and
(\ref{eq2}) was performed in the domain $(2\pi )^{3}$ with periodic
boundary conditions along all three coordinates. Integration over time
was performed by the explicit Runge-Kutta method of the fourth order of accuracy with
a step of $dt=5\cdot 10^{-3}$. The equations were integrated over spatial
coordinates by pseudo spectral methods with a total number
of harmonics $N^{3}=512^{3}$. To suppress the aliasing effect, a
filter was used that zeroed out higher harmonics with a wave number higher than $k_{a}\geq
N/3$. The operator $\gamma _{k}$ responsible for dissipation and the term responsible for pumping
the system $\mathcal{F}(\mathbf{k},t)$ were defined in Fourier space as:
\begin{eqnarray*}
\gamma _{k} &=&0,\quad k\leq k_{d}, \\
\gamma _{k} &=&\gamma _{0}(k-k_d)^2,\quad k>k_{d}, \\
\mathcal{F}(\mathbf{k},t) &=&F(k)\cdot \exp [iR(\mathbf{k},t)], \\
F(k) &=&F_{0}\cdot \exp [-\lambda_0^4 (k-k_{1})^{4}],\quad k\leq k_{2}, \\
F(k) &=&0,\quad k>k_{2}.
\end{eqnarray*}
Here $R(\mathbf{k,}t)$ are random numbers uniformly distributed in
the interval $[0,2\pi ]$, $\gamma _{0}$ and
$F_{0}$ are constants.

In this paper, three series of calculations were performed for different
regimes of acoustic turbulence. The first series of calculations was carried out
to demonstrate the possibility of realization of a weakly turbulent regime
of sound turbulence with a sufficiently small value of the dispersion parameter.
In the second series of calculations, a completely
non-dispersive regime $a=0$ was investigated, but with the same level of
nonlinearity. The third numerical experiment was also carried out for the non-dispersive  wave regime,
 but with a sufficiently high level of energy pumping to show the transition of the system to a strongly nonlinear motion regime.

For the first experiment, the dispersion length $a$ was equal to
$2.5\cdot10^{-3}$. In this case, the maximum dispersion addition at the
end of the inertial interval, $k=k_{d}$,
was $(k_{d}a)^{2}\approx 0.1\ll 1$. Other parameters were
chosen as follows: $k_{d}=110$, $k_{1}=3$, $k_2=6$,
$\lambda_0=0.64$, $\gamma_{0}=10^{-1}$, $F_{0}=1.25\cdot 10^{-3}$.
With this choice, the inertial interval was more than one decade. For
the second numerical experiment $a=0$, all other parameters remained unchanged, except for $\gamma_0=10^{-2}$.
The third series of calculations
was carried out for a strongly nonlinear non-dispersive regime. In this case,
the value of $F_0$ increased from $1.25\cdot
10^{-3}$ to $40\cdot10^{-3}$, i.e., by more than an order of magnitude. The chosen
value of $F_0$ is the maximum for our numerical model, since in this case
shock waves with a discontinuity width comparable to the grid step are formed.
At a large pumping amplitude, the bottleneck effect is observed, leading to the accumulation
of wave energy near the dissipation region $k_d$. To suppress this effect, the viscosity scale was chosen as $k_d=1$, i.e., energy dissipation
occurred in the entire range of wave numbers. Other parameters that were
changed in this case: $dt=2.5\cdot10^{-3}$, $\gamma_0=10^{-4}$.

\section{Simulation results}

\begin{figure*}[h!]
\centering
\includegraphics[width=4in]{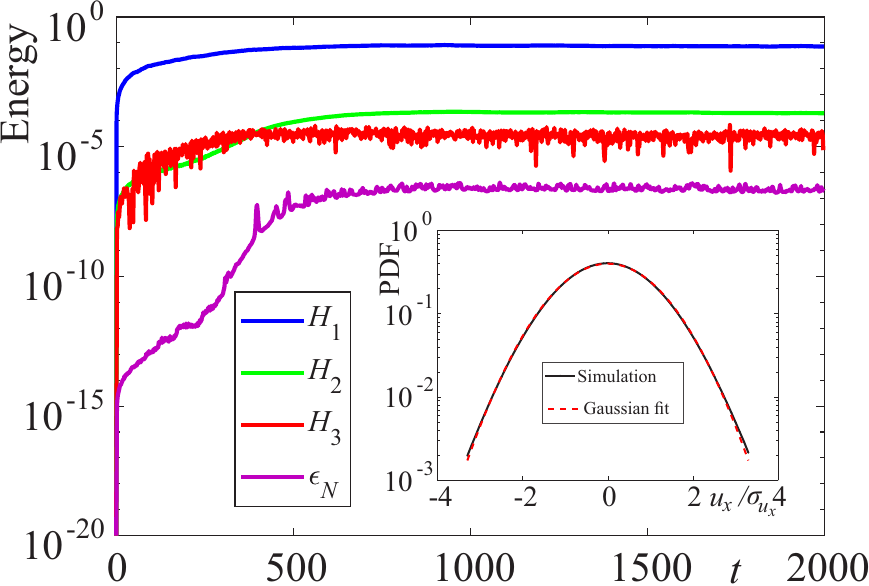}  \caption{Energy
contributions (\protect\ref{ham1}) and energy flux (\protect\ref{Ds})
are shown as a function of time for $a=2.5\cdot 10^{-3}$.
The inset shows the probability density function (PDF)
for $u_{x}$ measured relative to the standard
deviation $\protect\sigma _{u_{x}}$. The red dashed lines
correspond to a Gaussian distribution.}
\label{fig1}
\end{figure*}

\begin{figure*}[h!]
\centering
\includegraphics[width=4in]{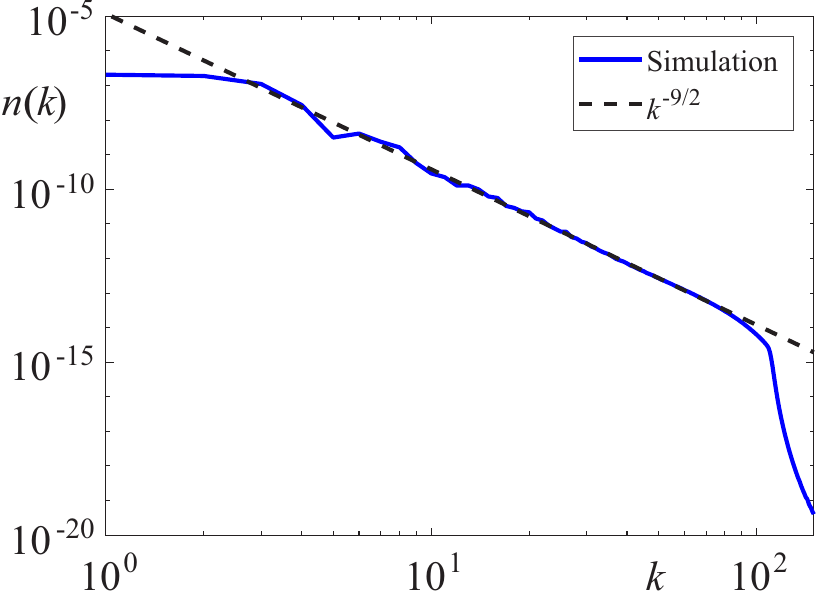}  \caption{Spatial
spectrum of wave action $n(k)$, calculated in the steady state, for the regime of weak
dispersion $a=2.5\cdot 10^{-3}$. The black dotted line
corresponds to the Zakharov-Sagdeev spectrum (\protect\ref{ZS}).}
\label{fig2}
\end{figure*}

Let us first present the results of modeling the weakly dispersive
regime of acoustic turbulence. To do this, in equations (\ref{eq1})
and (\ref{eq2}) we set $a=2.5\cdot10^{-3}$. Fig.~\ref{fig1} shows the
time dependence of the energy contributions to the Hamiltonian \eqref{ham1}. It is evident that the system rather quickly (by
time $t\approx 500$) passes to a stationary
chaotic evolution regime. It is important that the dispersion
contribution $H_{2}$ and the energy of nonlinear
interaction $H_{3}$ are small compared to the
term $H_{1}$, corresponding to the energy of linear non-dispersive
waves. In this case, the dispersion contribution of $H_2$ is almost an order of magnitude greater than the nonlinear energy of $H_3$, indicating the realization of a weakly nonlinear evolution regime. The inset to Fig.~\ref{fig1} shows the probability density function (PDF) measured in the quasi-stationary state for the derivative of $u_x$. As can be seen from the figure, the probability density is very close to the normal Gaussian distribution, which is valid for random uncorrelated signals. The turbulence spectrum averaged over angles in Fourier space in terms of the wave action $n(k)$ is shown in Fig.~\ref{fig2}. Indeed, one can see that the calculated spectrum exhibits a power-law distribution with an exponent close to the Zakharov-Sagdeev spectrum: $n_k\propto k^{-9/2}$.

\begin{figure*}[h!]
\centering
\includegraphics[width=4in]{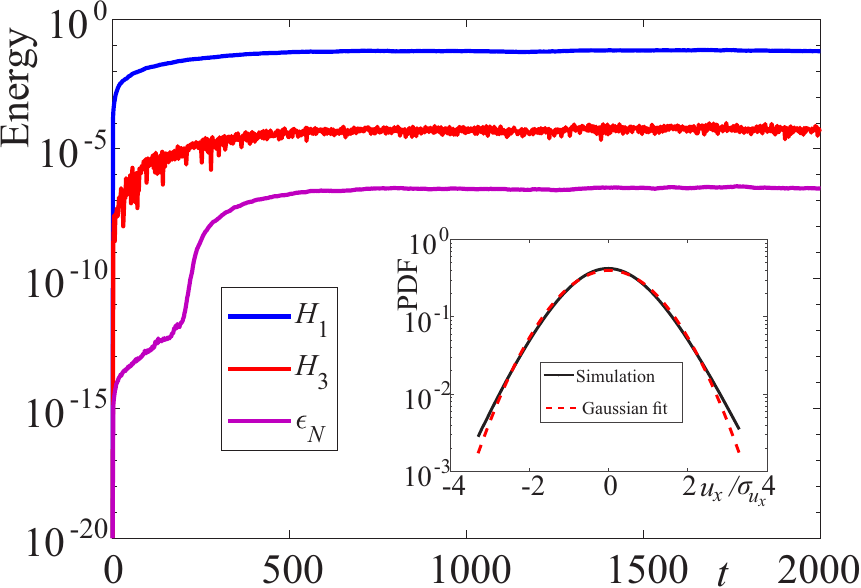}  \caption{Energy
contributions (\protect\ref{ham1}) and energy flux (\protect\ref{Ds})
are shown as a function of time for $a=0$. The inset shows the
probability density function (PDF) for $u_{x}
$ measured relative to the standard deviation
$\protect\sigma _{u_{x}}$. The red dotted lines correspond to a
Gaussian distribution.}
\label{fig3}
\end{figure*}

\begin{figure*}[h!]
\centering
\includegraphics[width=4in]{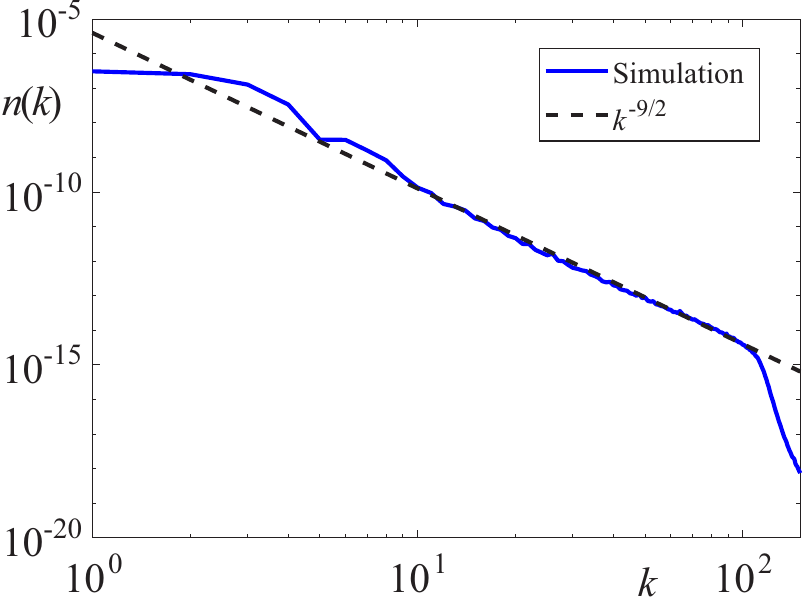}  \caption{Spatial
spectrum of wave action $n(k)$, calculated in the steady state, for the
dispersionless regime $a=0$. The black dotted line
corresponds to the Zakharov-Sagdeev spectrum (\protect\ref{ZS}).}
\label{fig4}
\end{figure*}

\begin{figure*}[h!]
\centering
\includegraphics[width=5.75in]{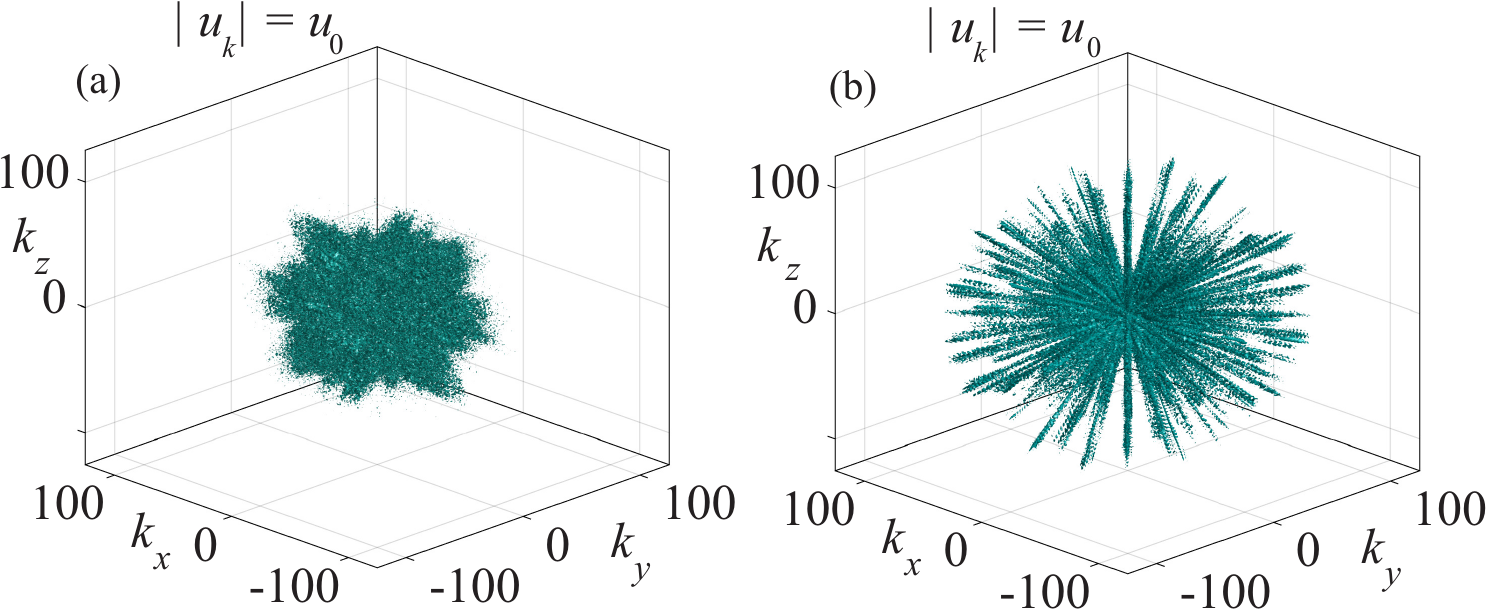}  \caption{Isosurfaces
of the Fourier spectrum $|u_k|=5\cdot10^{-6}$ (a)
for $a=2.5\cdot 10^{-3}$ and (b) $a=0$.}
\label{fig5}
\end{figure*}

The second numerical experiment was aimed at testing the possible
realization of the weak turbulence regime in the complete absence of dispersion,
i.e., at $a=0$. Fig.~\ref{fig3} shows that the system
of non-dispersive acoustic waves reaches a stationary
state at approximately the same time ($t\approx500$) as in the case of weak
dispersion. The intensity
of energy pumping $F_0$ was the same for both numerical experiments,
$F_0=1.25\cdot 10^{-3}$. The term responsible for the nonlinear
interaction
$H_3$ turns out to be three orders of magnitude smaller than the linear energy
$H_1$ in this regime.
The corresponding probability density for the quantity $u_x$,
shown in the inset to Fig.~\ref{fig3}, is close to the Gaussian
distribution, as in the weakly dispersive
case. Fig.~\ref{fig4} shows the turbulence spectrum for the non-dispersive regime $a=0$. The calculated wave action spectrum
is in very good agreement with the spectrum
of Zakharov-Sagdeev $k^{-9/2}$ (\ref{ZS}), written in terms of $n_k$.
Thus, upon transition to the non-dispersive regime, the system remains in a weakly turbulent state. The question arises: what is the mechanism for the
realization of weak acoustic turbulence in the absence of wave dispersion? To answer this question, Fig.~\ref{fig5}
shows the Fourier isosurfaces of the spectra for two regimes. It can be seen that in the non-dispersive regime, see Fig.~\ref{fig5}~(b), structures with a large
number of jets in the form of narrow cones arise in the distribution of turbulent pulsations. The emergence of such
structures is the result of resonant wave
interactions (\ref{3wave}). In the weakly dispersive regime, jets
are observed only at very small $k$, close to the pumping region,
when dispersion can be neglected. In the region of large wave
numbers, jets are smoothed out, and the spectrum distribution
correspondingly tends to be isotropic, shown in
Fig.~\ref{fig5}~(a). In the absence of dispersion, such smoothing does not
occur and the spectrum is a discrete set of narrow jets.

\begin{figure*}[h!]
\centering
\includegraphics[width=5.75in]{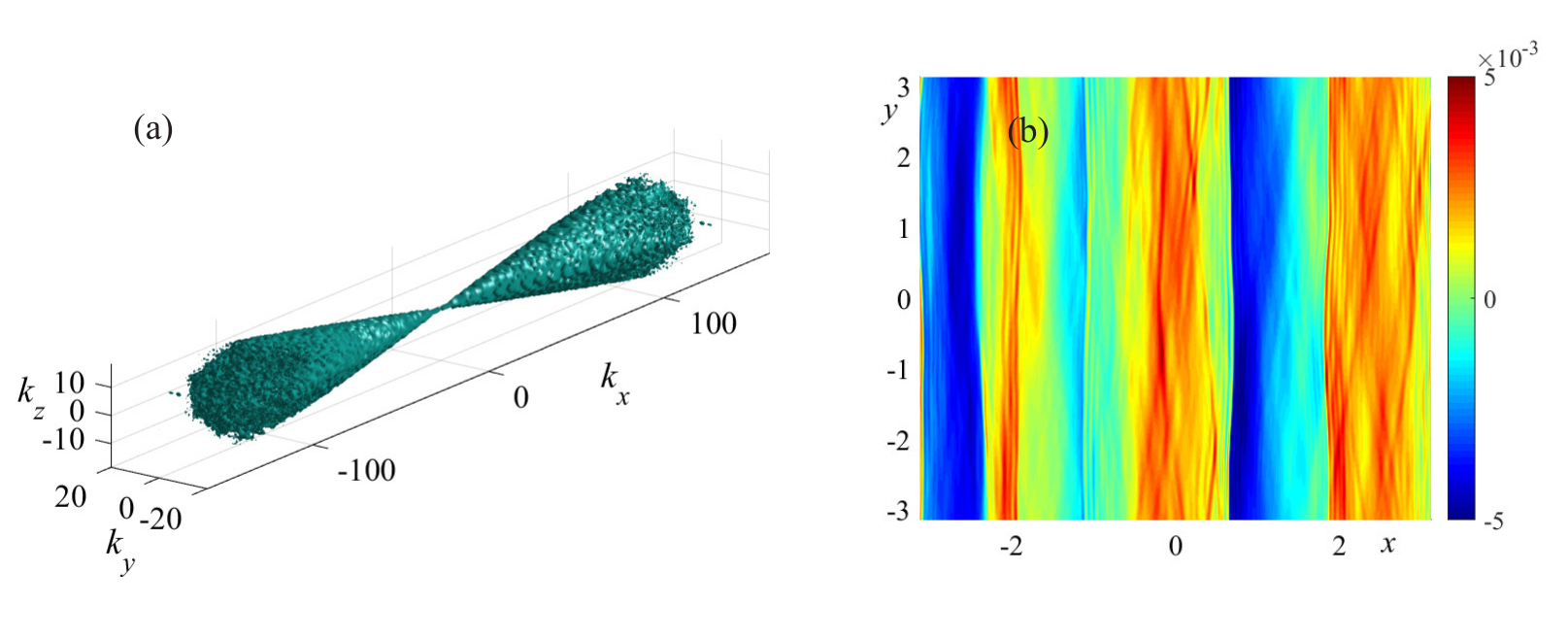}  \caption{(a) Isosurface of the Fourier spectrum $|u(\mathbf{k})|=10^{-7}$ for a single jet
inside the angle $\protect\pi/10$.
(b) The inverse Fourier transform of the jet in the
$\{x,y\}$ plane at time $t = 2000$, corresponding to the quasi-stationary state, is shown.}
\label{fig6}
\end{figure*}

To clarify the mechanism of weak turbulence  in the non-dispersive regime,
a separate jet directed along the $x$ axis was isolated using filtering in the Fourier space.
 For such a narrow cone (jet), the dispersion relation \eqref{disp1} ($a=0$) can be expanded in a series with respect to the small
parameter $\Omega _{0}=k_{\perp}/k_x$ (the cone angle):
\begin{equation*}
\omega _{k}=|\mathbf{k}|\approx k_{x}\left( 1+\Omega _{0}^2/2\right),
\end{equation*}
so that $H_{1}$ is the sum of two terms
\begin{equation*}
H_{1}=\int \omega _{k}n(k)d\mathbf{k}\approx \int {[k_{x}+\Omega
_{0}^{2}k_{x}/2]n(k)d\mathbf{k}}=H_{\parallel }+H_{\perp },
\end{equation*}
where $H_{\parallel }$ is the energy of the acoustic beam propagating
along the $x$ direction, and $H_{\perp}$ determines its
diffraction energy. To calculate these contributions, we filtered
the function $u(\mathbf{k})$ in a narrow cone along $k_{x}$ with an
apex angle of $\Omega_0=\pi/10$. The result of such filtering is shown
in Fig.~\ref{fig6}. Thus, the energy contributions have
the ratio $H_{\perp }/H_{\parallel }\approx 0.05$. Direct calculation
gives $H_{\perp }\approx 4.3\cdot 10^{-5}$, while the energy
of nonlinear interaction for the selected cone is estimated as
$H_{3}\approx 3.7\cdot 10^{-7}$. Thus, for an individual jet,
the diffraction energy turned out to be two orders of magnitude greater than the energy of nonlinear interaction,
which explains the occurrence of weak acoustic turbulence.

\begin{figure*}[h!]
\centering
\includegraphics[width=5.75in]{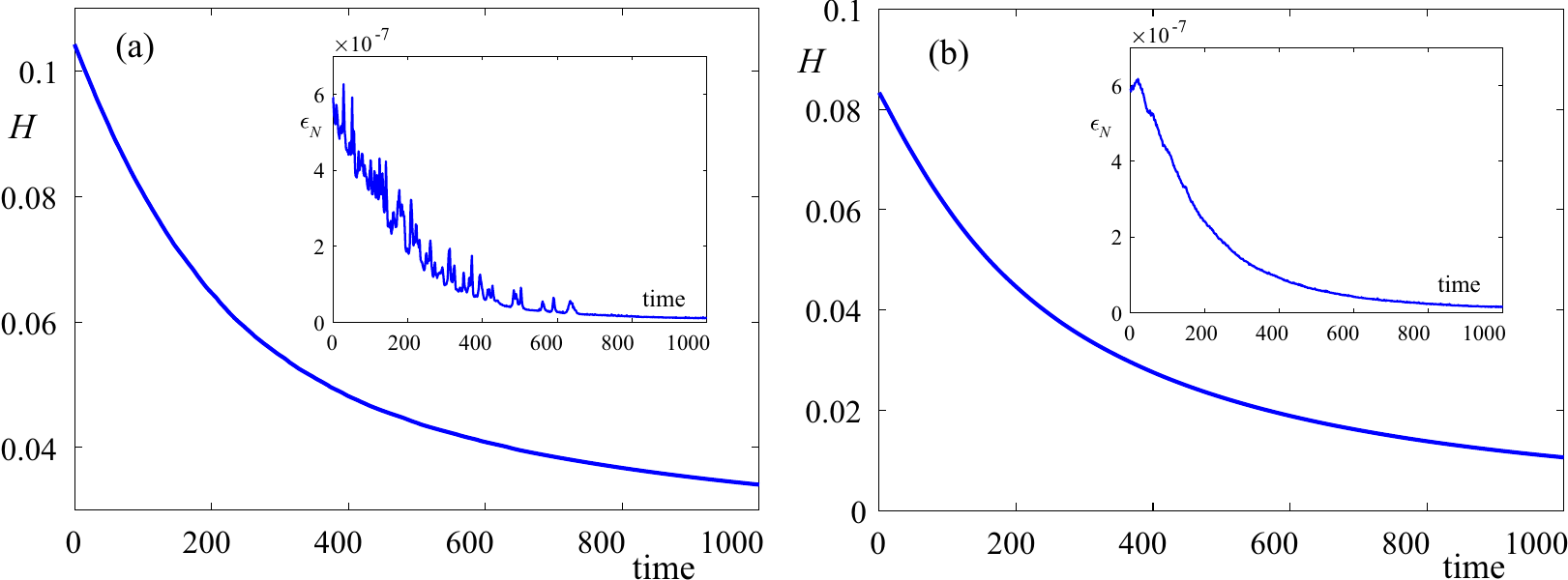}  \caption{Evolution of the total energy for (a) $a=2.5\cdot 10^{-3}$ and (b) $a=0$. The insets show the time dependence of the energy dissipation flux.}
\label{fig7}
\end{figure*}

\begin{figure*}[h!]
\centering
\includegraphics[width=3.5in]{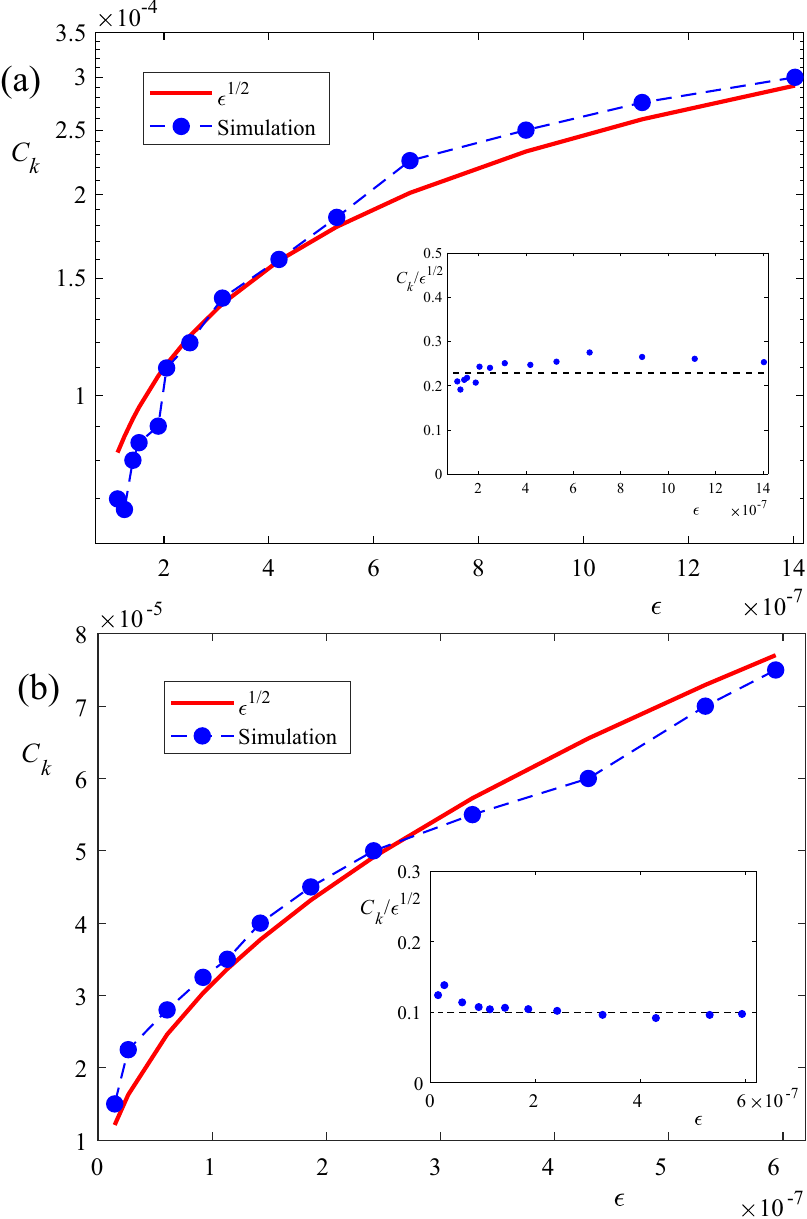}
\caption{The $C_k$ coefficients are shown as functions of the measured
energy flux for (a) $a=2.5\cdot 10^{-3}$ and
(b) $a=0$. The red solid lines show the square root
$\protect\epsilon^{1/2}$ dependence. The insets show the compensated
$C_k/\protect\epsilon^{1/2}$ (Kolmogorov-Zakharov's constant) values.}
\label{fig8}
\end{figure*}

Thus, weak acoustic turbulence can be realized
both in the weak dispersion regime and in the completely non dispersion case.
The criterion for the realization of weak turbulence in both regimes is in some sense the same: the effects of dispersion or diffraction (in the case of
$a=0$) should prevail over the nonlinearity, which is responsible for the
breaking of acoustic waves. An important
characteristic of the weakly turbulent regime of acoustic turbulence
is the Kolmogorov-Zakharov constant $C_{KZ}$, which is part of the
spectrum (\ref{ZS}). For its numerical evaluation, we carried out two series of calculations
for weakly dispersion and non dispersion regimes of wave
turbulence. The numerical experiments simulated decaying
turbulence at $F_0=0$. The initial conditions were taken at the time $t=2000$, corresponding to the quasi-stationary state.
The calculations were performed on the time interval $T=1000$. As can be seen from
Fig.~\ref{fig7}, during this period of time the total energy decreases by several times, and the dissipation flow almost stops, i.e., the system
becomes almost
linear. In the mode of decaying turbulence, the evolution of the turbulence spectra is self-similar with the preservation of the
exponent $-9/2$, but the coefficient before the power decreases with time
as a result of energy dissipation. To estimate the Kolmogorov-Zakharov constant, we present the calculated energy
spectra in the form $E(k)=C_k k^{-3/2}$, where the value of $C_k$
was calculated for two modes depending on the energy
flow $\epsilon$. The measured functions agree well with the
analytical dependence $C_k\propto\epsilon^{1/2}$, both for
weakly dispersive and for non-dispersive
regimes, see Fig.~\ref{fig8}. The insets to Fig.~\ref{fig8}
allow us to directly estimate the values of the constants for two
regimes: $C_{KZ}\approx 0.24$
(weakly dispersive) and $C_{KZ}\approx 0.1$ (non-dispersive).
Thus, the numerical estimate for the Kolmogorov-Zakharov constant in the weakly
dispersive
regime agrees much better with the exact
value $C_{KZ}\approx 0.22$. In the non-dispersive
regime, the value of $C_{KZ}$ differs from (\ref{KZaa}) by approximately
a factor of two. Apparently, such a difference arises as a result of the anisotropic distribution of energy
in the Fourier space, observed in Fig.~\ref{fig5}~(b). In the presence of dispersion, the turbulence spectrum
tends to an isotropic distribution, providing a value of the Kolmogorov-Zakharov constant closer to the theoretical value (\ref{KZaa}).

\begin{figure*}[h!]
\centering
\includegraphics[width=4in]{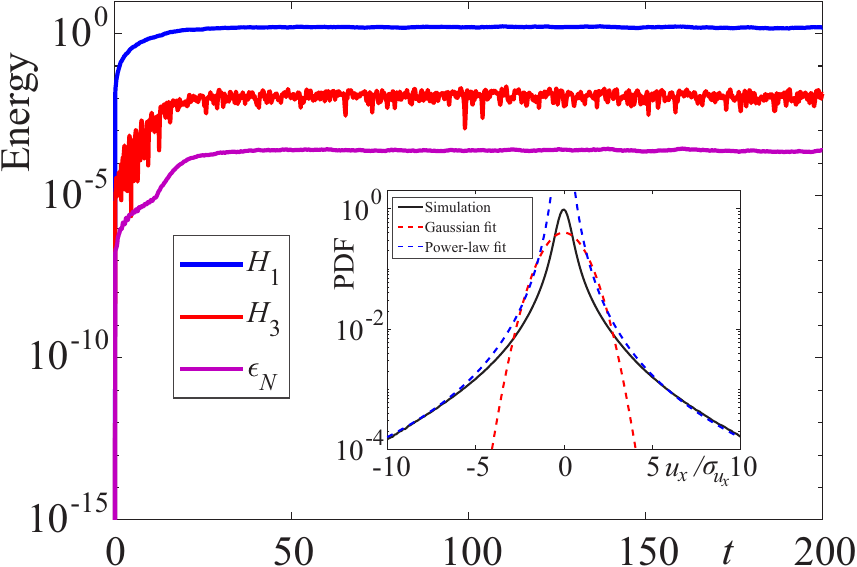}  \caption{Energy
contributions (\protect\ref{ham1}) and energy flux (\protect\ref{Ds})
are shown as a function of time for $a=2.5\cdot 10^{-3}$.
The inset shows the probability density function (PDF)
for $u_{x}$ measured relative to the standard
deviation $\protect\sigma _{u_{x}}$. The red dotted lines
correspond to a Gaussian distribution, the blue dotted lines to a power
law: $|u_{x}/\protect\sigma _{u_{x}}|^{-7/2}$.}
\label{fig9}
\end{figure*}

\begin{figure*}[h!]
\centering
\includegraphics[width=4in]{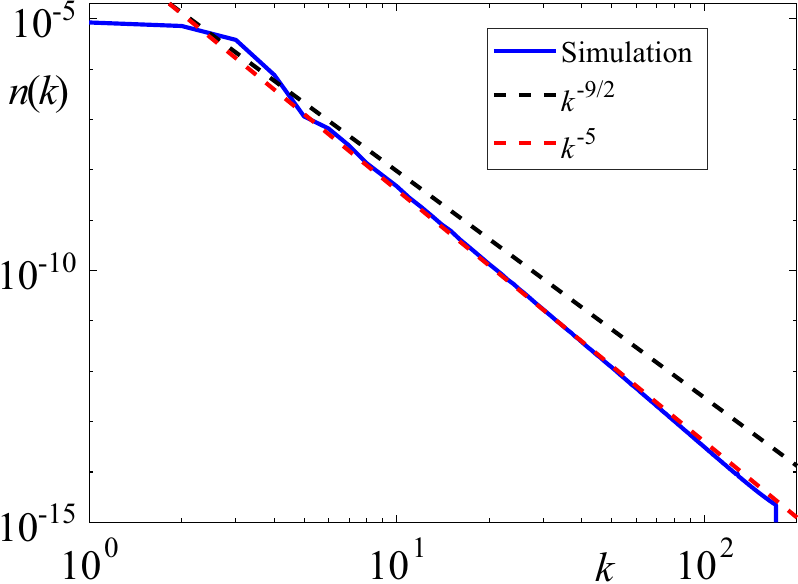}  \caption{Spatial
spectrum of wave action $n(k)$, calculated in the steady state, for the
non-dispersive regime $a=0$. The black dotted line
corresponds to the Zakharov-Sagdeev  spectrum (\protect\ref{ZS}), the red
dotted line is the Kadomtsev-Petviashvili  spectrum (\protect\ref{KP}).}
\label{fig10}
\end{figure*}

\begin{figure*}[h!]
\centering
\includegraphics[width=4in]{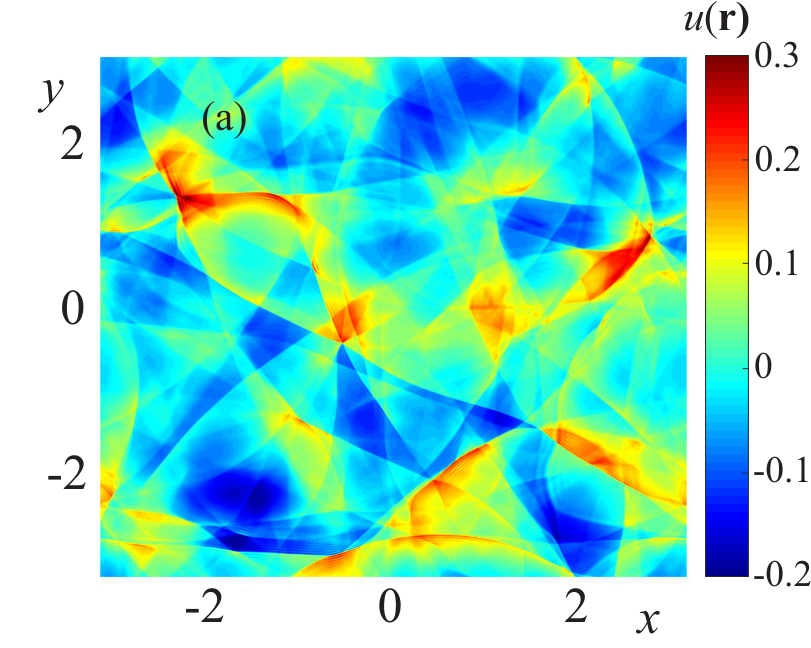}  \caption{Distribution
of the quantity $u(\mathbf{r})$ in the plane $z=0$ in the regime of strong acoustic
turbulence.}
\label{fig11}
\end{figure*}

Let us now consider the last third series of calculations. The goal of this part of the work was to simulate
 the non-dispersive regime of wave turbulence, but with an increase in the level of nonlinearity of the system.
 For this purpose, the pumping amplitude $F_{0}$ was significantly increased to $40\cdot10^{-3}$,
i.e., more than an order of magnitude greater compared to the two previous
numerical experiments. For such conditions we observed a very
fast transition to the steady state, in times $t\approx 50
$, see Fig.~\ref{fig9}. In the inset to Fig.~\ref{fig9} one can
see that the probability density function is very different from the Gaussian
distribution with strongly elongated non-Gaussian tails, indicating the presence of
extreme
events (shock waves). Note that for the one-dimensional
Burgers equation with weak viscosity and random pumping, the tails of the probability density function for the velocity gradient decay according to a
power law with exponent $-7/2$, see details
\cite{pdf1,pdf2,pdf3, pdf4,pdf5,pdf6}. In the inset in Fig.~\ref{fig9}, it can be seen that the
calculated probability density for $u_x$ is described with high accuracy by the same power-law dependence: $|u_x|^{-7/2}$.
Thus, Fig.~\ref{fig9} indicates a high level of intermittency in the system.
The turbulence spectrum for a strongly nonlinear regime is shown in Fig.~\ref{fig10}: the exponent for $n(k)$
is close to $-5$, which in terms of wave action
corresponds to the Kadomtsev-Petviashvili \eqref{KP} spectrum. Fig.~\ref{fig11} shows the spatial distribution
of $u(\mathbf{r})$ in the $z=0$ plane. It can be clearly seen that a large number of discontinuities have formed in the
distribution of $u(\mathbf{r})$. In fact, such discontinuities are a set of shock waves,
propagating at different angles to each other. Thus,
at a large pumping amplitude, a transition to a
state of strong acoustic turbulence is indeed observed, described by the Kadomtsev-Petviashvili spectrum (\ref{KP}) and representing an
ensemble of shock waves chaotically propagating in space due to the random nature of the system pumping.

\section{Conclusion}

In this, review we have presented the research results on three-dimensional sound turbulence. This study includes both analytical investigations going back to the works of Zakharov and Sagdeev on weak wave turbulence \cite{zakh65, zs-70} and to the work of Kadomtsev and Petviashvili \cite{KP} on strong sound turbulence. The main results of this review were recently published by the authors in \cite{KochurinKuznetsov1,KochurinKuznetsov2}
Based on the direct numerical simulation technique, it was found that the transition from a weakly turbulent regime to strong turbulence
is determined by the ratio of linear (dispersion or diffraction) effects to nonlinearity.
A numerical experiment confirmed the existence of weak sound turbulence with the Zakharov-Sagdeev spectrum ($~ k^{-3/2}$) for both weak positive dispersion waves and non-dispersive sound waves. The calculations revealed that the spectrum of weak sound turbulence is strongly anisotropic, especially in the region of small $k$, where the distribution in $k$-space consists of a set of jets. In the weak positive dispersion regime, the jets broaden with increasing $k$ and the spectrum becomes almost isotropic. For non-dispersive sound waves, the distribution in Fourier space is a finite number of cone-shaped jets, for each of which diffraction across the jets prevents the breaking of sound waves, realizing the weak sound turbulence regime. In other words, diffraction plays the same role in this case as dispersion. It should be noted that for weak nonlinearity each jet in the non-dispersive limit $a=0$ can be described by the Khokhlov-Zabolotskaya equation \cite{ZK}, which describes the competition between the nonlinear interaction and beam diffraction. In our numerical experiments, when the weakly turbulent Zakharov-Sagdeev spectrum is realized, diffraction effects dominate. In the strongly nonlinear turbulent regime, the jet structure disappears and the spectrum turns out to be close to isotropic due to shock waves propagating at different angles (see Fig. 11). The formation of shock waves in sound beams described by the Khokhlov-Zabolotskaya equation was also observed in numerical simulations (see the review \cite{rudenko} and references therein). However, the distribution remained quasi-one-dimensional.

In weak turbulence, the probability density distribution of wave field gradients turns out to be close to Gaussian. In the regime of strong nonlinearity for sound waves without dispersion, the main effects are the breaking and formation of shock waves. This is the reason for the formation of Kadomtsev-Petviashvili spectrum. In this case, power-law tails for large deviations of the gradient appear in the probability density distribution (PDF), similar to those predicted by the Burgers turbulence theory with an index close to $-7/2$. Such tails indicate the intermittency of Kadomtsev-Petviashvili turbulence. Thus, the sound turbulence is transformed into an ensemble of random shock waves, and is described with high accuracy by the Kadomtsev-Petviashvili spectrum \eqref{KP}.
The calculations are based on pseudo-spectral methods of direct numerical modeling of acoustic turbulence. Simulations were carried out in a wide range of control parameters and allowed to establish criteria for the formation of a weak or strong turbulent regime. Thus, the Zakharov-Sagdeev \eqref{ZS} and Kadomtsev-Petviashvili \eqref{KP} spectra correspond to different limiting cases of weakly and strongly nonlinear regimes of three-dimensional acoustic
turbulence with respect to linear wave effects.

In conclusion, we would like to discuss when power-law ($~ k^{-3/2}$)  turbulent spectra of sound-type waves arise. As noted above, the numerical experiment \cite{Galtier1} for turbulence of fast magnetosonic waves in plasma with small $\beta$ demonstrates the Zakharov-Sagdeev spectrum, since waves of this type have the sound dispersion law $\omega=kV_A$ (here $V_A$ is the Alfv\'en velocity) in the frequency range below the ion cyclotron  $\omega_{ci}$. The spectrum of slow magnetosonic waves (in plasma with small $\beta$ they can be considered as magnetized ion sound) demonstrates the same dependence on $k$ as the Zakharov-Sagdeev spectrum, corresponding to a constant energy flow into the region of short waves, despite strong anisotropy. For small $\beta$, the same dependence on the modulus $k$ is shown by the spectra of weak turbulence of fast magnetosonic and Alfv\' en waves interacting with slow magnetosonic waves \cite{KuznetsovMHD}. Recent experiments (see the papers \cite{Magnetic1,Magnetic} and references therein) on observing the spectra of magnetic field fluctuations in the solar wind near the Sun have shown dependencies close to the spectrum $~\omega^{-3/2}$, where the parameter $\beta$ is small, of the order of $0.1$. The Kadomtsev-Petviashvili spectrum, as shown by numerical experiments \cite{Popova} with applications to astrophysics, is often encountered in MHD turbulence, as is the spectrum $~ k^{-3/2}$. It is important that the anisotropy of the spectra due to the magnetic field does not affect the dependence of the spectra on the absolute value of $k$. For weak MHD turbulence, this behavior follows from simple dimensional estimates.

\section{Acknowledgements}

The authors express their sincere gratitude to V.E.~Zakharov for fruitful discussions at the early stage of this work. We are grateful to S.N.~Gurbatov, who drew our attention to the articles on Burgers turbulence \cite{pdf1}-\cite{pdf6}, and to R.Z.~Sagdeev, who pointed out the numerical modeling of MHD wave turbulence in astrophysics. The work was supported by the Russian Science Foundation (grant: 19-72-30028)

\end{document}